\documentclass[sigconf]{acmart}

\AtBeginDocument{%
  \providecommand\BibTeX{{%
    \normalfont B\kern-0.5em{\scshape i\kern-0.25em b}\kern-0.8em\TeX}}}

\usepackage{eiffel}
\usepackage{wrapfig}
\usepackage{soul}
\usepackage{enumitem}
\setitemize{leftmargin=*}

\usepackage{etoolbox} 
\usepackage{xstring}  

\newcommand{\e}[1]{\mbox{\lstinline[basicstyle=\normalsize]|#1|}}

\newcommand{\foote}[1]{\mbox{\color[HTML]{222277}\footnotesize\bfseries\ttfamily}}

\setcopyright{none}
\renewcommand\footnotetextcopyrightpermission[1]{} 
\begin{document}

\title{Do AI models help produce verified bug fixes?}

\author{Li Huang,  Ilgiz Mustafin, Marco Piccioni, Alessandro Schena, Reto Weber, Bertrand Meyer}
\email{<first.last@constructor.org>}

\affiliation{%
 \institution{Constructor Institute of Technology}
 \streetaddress{Rheinweg 9}
 \city{Schaffhausen}
 \country{Switzerland}
 \postcode{8212}
 }


\begin{abstract}
Among areas of software engineering where AI techniques --- particularly, Large Language Models --- seem poised to yield dramatic improvements, an  attractive candidate is Automatic Program Repair (APR), the production of satisfactory corrections to software bugs. Does this expectation materialize in practice? How do we find out, making sure that proposed corrections actually work? If programmers have access to LLMs,  how do they actually use them to complement their own skills?

To answer these questions, we took advantage of the availability of a program-proving environment, which formally determines the correctness of proposed fixes, to conduct a study of program debugging with two randomly assigned groups of programmers, one with access to LLMs and the other without, both validating their answers through the proof tools. The methodology relied on a division into general research questions (Goals in the Goal-Query-Metric approach), specific elements admitting specific answers (Queries), and measurements supporting these answers (Metrics). While applied so far to a limited sample size, the results are a first step towards delineating a proper role for AI and LLMs in providing guaranteed-correct fixes to program bugs.

These results caused surprise as compared to what one might expect from the use of AI for debugging and APR. The contributions also include: a detailed methodology for experiments in the use of LLMs for debugging, which other projects can reuse; a fine-grain analysis of programmer behavior, made possible by the use of full-session recording; a definition of patterns of use of LLMs, with 7 distinct categories; and validated advice for getting the best of LLMs for debugging and Automatic Program Repair.

\end {abstract}






\maketitle
\vspace{1.5cm}
\noindent \textit{Note}: This article was entirely written by the authors. No part of it was generated by an LLM or other automatic tool.
\section{Introduction} \label{introduction}

As the AI wave --- particularly the LLM wave --- upends ever more areas of industry and knowledge, software engineering is on the front line. Predictions that AI techniques will replace programmers run wild. While  opinions differ on whether it will truly be a replacement or just a complement, some of the transfer has already happened; as one example, the Microsoft CEO  publicly stated that as of April 2025 30\% of the company's code is ``written by AI'' \cite{Nadella2025}.

A widely popular practice among developers is ``vibe coding'', which uses AI to assist programming, often producing large chunks of the code. Software engineering is, however, more than coding. What about vibe designing, vibe specifying, vibe testing, vibe program proving or --- the focus of this article --- vibe debugging?

Seeking help on the verification side, including debugging, is a natural move. In software engineering as in other disciplines, the succession of reactions when one starts using LLM is often a seesaw: alternated "\textit{Wow, see what it just did!}" and "\textit{Wow, see how wrong it is ---  hallucinating again!}" reactions. LLMs have been shown to produce fundamentally wrong solutions with airs of absolute assurance \cite{meyer2023ai}. Hence the idea of complementing the impressive but occasionally wayward creativity of LLMs with the boring but reassuring rigor of formal verification techniques supported by mechanical proof tools. That is what the study reported in this article does. We presented a group of programmers with:
\begin{itemize}
    \item A set of buggy programs.
    \item A request to find and correct the bugs.
    \item Access to a set of state-of-the-art LLMs --- or, for the control group, no such help.
    \item A formal verification environment (a mathematics-based program prover) to validate the fixes.
\end{itemize}

\noindent We not only collected the results but video-taped the entire session to try to gain detailed insights on \textit{whether} LLMs help towards finding bug fixes, and, if they do, \textit{how} programmers take advantage of them. These are our two basic research questions, further detailed in section \ref{questions}.

This study is a contribution to Automatic Program Repair (APR), a thriving field of research in the past two decades \cite{monperrus2018automatic, monperrus2018living}. A burning question in that field is the extent to which it can use AI for proposing corrections (``fixes'') to buggy programs. We provide elements of answer with a special feature: we take advantage of technology that can \textit{formally verify} proposed fixes. In much APR work, the final step of the APR process, checking that the proposed fixes do indeed work and remove the bug, relies on testing. It is the weakest link in the traditional APR chain since it assumes that: (1) programmers produce a set of test cases, a tedious task; (2) these test cases are different from those that led to identification of the bug (otherwise, they risk overfitting: fixing the program's behavior only for the identified cases, without correcting the actual fault); (3) the test cases, even without blatant overfitting, are significant. A test is only the program's result on one set of inputs, not a specification. In contrast, the present work validates fixes through an automatic program prover. The downside is that programmers must have written a precise specification in the form of \textit{contracts} \cite{meyer1992applying}, but then the proof is an actual mathematical guarantee of correctness. So the present work is not just evaluating the use of AI for generating bug fixes, but its use for generating \textit{verified} bug fixes.

The analysis applies to the research questions the GQM method of empirical software engineering, Goal-Question-Metric \cite{GQM}.

Section \ref {terminology}  defines basic APR terminologies. Section \ref{questions} presents the research questions (the goals) as well as the queries and metrics. Section \ref{related} presents related work. Section \ref{tools} introduces the tools used for the study, and Section \ref{setup} the setup (participants, tasks, mode of recording). Section \ref{threats} discusses limitations and threats to validity. Section \ref{analysis} presents the measured results, complemented in section \ref{observations} by more qualitative observations resulting from observing programmers' behaviors: practices that hamper fruitful  usage of LLMs (``antipatterns'', \ref{antipatterns}), LLM user ``personalities'' (\ref{patterns}, advice for LLM usage (\ref{advice}.). Section \ref{conclusions} is the conclusion.

\vspace{-0.3cm}
\section{Terminology}\label{terminology}

A malfunction in a program is called a \textit{failure}. The underlying deficiency is a \textit{fault}, also informally called a \textit{bug}. (One fault can result in several different failures in various executions of the program.) A proposed remedy to the fault, hoped to remove the fault, is a \textit{correction}, also called a \textit{fix}, a \textit{repair} or a \textit{patch}.

A fix is \textit{valid} if it leads to a valid program (syntactically and type-wise correct, in other words able to be compiled and run) and removes the failure or failures that were observed. A valid fix is \textit{correct} if it actually removes the fault. The difference between ``valid'' and ``correct'' is related to the problem of \textit{overfitting}: if a number of failures were identified, a valid fix might remove these specific failures but still not address the underlying problem. (As an extreme but worthless solution, the fix might just add a conditional instruction that produces the expected results in the identified cases, and for all others proceeds as before.)

Most APR approaches are \textbf{test-based}: not only are failures identified as a result of test runs not producing the expected effect, but the checking of a valid fix for correctness also uses a set of test cases (a different one, to avoid overfitting). This notion of correctness is not absolute. For a firmer assessment of correctness, one needs a \textbf{proof-based} approach, assuming the presence of a formal specification of each program element's function, and a tool to validate an implementation against it. In the tool stack described in section \ref{tools} programs are written in Eiffel and include a specification in the form of ``contracts'', and the AutoProof program prover can check an implementation --- in particular, an implementation incorporating a fix --- against the contracts, through mathematical techniques and tools. (That notion of correctness is still not total: it only expresses that the program is correct with respect to the given contracts, ignoring any other aspects; and it is dependent on the correctness of the proof tools themselves. But it is rigorously defined and much more credible than a test-based view.)

\section{Research questions}\label{questions}

Automatic Program Repair (APR) has been around in the past decade and a half, resulting in a wide array of techniques and tools, still mostly at the stage of prototypes although some elements have found their way into IDEs (development environments used by programmers). The widely accepted typical APR cycle \cite{le2011genprog} includes four phases: fault localization, fix generation, fix validation (including for correctness) and fix selection (when several fixes are being proposed).

Some fix generation techniques are very sophisticated, but it is a normal reaction for a researcher in the field to compare the results of the latest heuristics with what an LLM such as ChatGPT would produce, based on modern AI approaches (machine learning and generative techniques). The first results are often impressive, suggesting that these approaches can be effective at fix suggestion. Beyond such an individual experiences, there has not been (to our knowledge) any systematic evaluation of how good LLMs are at generating valid and correct fixes or --- more realistically, since in the current state of technology we cannot rule out the presence of a human in the loop --- at helping programmers obtain such fixes. This question is the focus of the study.

More specifically, we set out to obtain elements of answer the following two research questions, which for mnemonics we call W for ``whether'' and H for ``How'' (instead of the usual RQ1 and RQ2):

\begin{itemize}
    \item W: In turning a buggy program into a correct one, is it fruitful to use an LLM? (``Whether''.)

    \item H: If programmers do use an LLM for debugging, what is an effective process? (``How''.)

\end{itemize}

The How question, H, assumes by default a positive answer to the Whether question, W, although even in the case of a negative answer, suggesting that LLMs do not help, it is still interesting to elicit  the ``how'' so as to understand where and why the LLMs fail the programmer.

The study and its analysis follow the GQM method, Goal-Query-Metric \cite{caldiera1994goal}, of performing software engineering studies. In this approach, we pursue some goals, defined as an effort to obtain answers to specified questions; here the research questions W and H are the goals. For that purpose, one should define \textit{queries}: specific, well-defined and concrete questions (often although not always Boolean), whose answers help address the underlying goals. To answer the queries, one defines \textit{metrics}: criteria that can be assessed by conducting a study, with a numerical value.

We defined the following queries and the corresponding metrics as follows. ``Q'' is for queries, ``M'' for metrics.

\emph{For goal W: In turning a buggy program into a correct one, is it fruitful to use an LLM?}

\begin{itemize}
    \item QW-1 Can programmers solve more debugging tasks with/without LLM?
          \begin{itemize}
              \item [] MW-1.a How many of the tasks are solved with LLM?
              \item [] MW-1.b How many of the tasks are solved without LLM?
          \end{itemize}

    \item QW-2 For tasks that programmers can solve with/without LLM, which takes longer?
          \begin{itemize}
              \item [] MW-2.a How long does each task solved with LLM take?
              \item [] MW-2.b How long does each task solved without LLM take?
          \end{itemize}
    
    \item QW-3 For tasks that programmers cannot solve with/without LLM, which takes longer?
          \begin{itemize} 
              \item [] MW-3.a How long does each task unsolved with LLM take?
              \item [] MW-3.b How long does each task unsolved without LLM take?
          \end{itemize}

    \item QW-4 Do submissions contain more incorrect features with/without LLM?
          \begin{itemize}
              \item [] MW-4.a How many tasks are incorrectly solved with LLM?
              \item [] MW-4.b How many tasks are incorrectly solved without LLM?
          \end{itemize}

    \item QW-5 Do LLMs help experienced programmers more/less than novice ones?
          \begin{itemize}
              \item [] MW-5.a How many problems do experienced programmers solve with/without LLM?
              \item [] MW-5.b How many problems do programmers with static verification experience  solve with/without LLM?
              \item [] MW-5.c How many problems do programmers with programming language experience solve with/without LLM? 
            \end{itemize}

    \item QW-6 Do LLMs help active programmers more/less than occasional ones?
          \begin{itemize}
              \item [] MW-6.a How many problems do active programmers solve with/without LLM?
              \item [] MW-6.b How many problems do programmers with active static verification practice solve with/without LLM?
              \item [] MW-6.c How many problems do active programmers in the given programming language solve with/without LLM?
          \end{itemize}
          
\end{itemize}

\textit{For goal H: If programmers do use an LLM for debugging, what is an effective process?}

\begin{itemize}
    \item QH-1 What are the categories of the prompts used by the programmers?
    \begin{itemize} 
              \item [] MH-1.a In using LLMs for debugging, what are the categories of prompts (determined empirically from the evidence and from previous published work)?
              \item [] MH-1.b For each task, how many categories of prompts do programmers use?
          \end{itemize}

    \item QH-2 What are the components of the prompts for solved/unsolved submission?
          \begin{itemize}
              \item [] MH-2.a How many prompts contain natural language?
              \item [] MH-2.b How many prompts contain code?
              \item [] MH-2.c How many prompts contain tool output?
              \item [] MH-2.d Is the task solved?
          \end{itemize}

    \item QH-3 How important is the phrasing of prompts in solving a bug?
          \begin{itemize}
              \item [] MH-3.a For each category of prompts, how many prompts lead to a solved task?
              \item [] MH-3.b How many prompt categories lead to a solved task?
          \end{itemize}

    \item QH-4 How do programmers interact with LLMs?
          \begin{itemize}
              \item [] MH-4.a How many prompts do programmers send before solving a task?
              \item [] MH-4.b How many prompts do programmers send for unsolved tasks?
          \end{itemize}

    \item QH-5 How do programmers use the outputs from LLMs?
          \begin{itemize}
              \item [] MH-5.a How many times do programmers copy-paste a fix produced by LLMs?
              \item [] MH-5.b How many times do programmers accept LLMs’ output as the final version?
          \end{itemize}

    \item QH-6 What is the effect of programmers’ experience\footnote{A more general version of this question would include \textit{age} in addition to experience. After giving it some consideraton we decided that it was nor essential to this study.} on LLM use?
          \begin{itemize}
              \item [] MH-6.a Do experienced programmers use more/fewer prompts?
              \item [] MH-6.b Do experienced programmers use LLM output more/less often?
          \end{itemize}
\end{itemize}

\noindent As a matter of methodology, the research questions as presented above (goals, queries and metrics) were defined as we were setting up the experiments, but before running them, and we did not modify them in any way afterwards, to avoid any a-posteriori confirmation bias (sometimes known as HARKing, for Hypothezing After the Results are Known \cite{kerr1998harking}).

\section{Related work} \label{related}




State-of-the-art large language models including ChatGPT, Claude, Gemini, Mistral and DeepSeek have demonstrated significant potential in supporting a variety of programming tasks, such as code generation \cite{barke2023grounded} and automatic program repair \cite{xia2024automated, wei2023copiloting}. Recent studies have further investigated their applications to a more advanced goal --- achieving formally verified programs. Such efforts include using LLMs to generate program specifications \cite{liu2024propertygpt} and synthesize code that conforms to specified requirements \cite{sun2024clover, brandfonbrener2024vermcts, teuber2025next}. In the context of fixing formally verified programs, Tihanyi et al. \cite{tihanyi2023new} established a workflow similar to the present study, combining LLMs with a formal verification tool: they first used Bounded Model Checking (BMC) to detect vulnerabilities and produces counterexamples; they fed the outputs from BMC to the LLM for generating fixes; the fixes  are then validated by BMC to confirm their correctness.

As LLMs become increasingly integrated into everyday software development workflows, providing an accurate estimation on their effectiveness across diverse programming tasks becomes crucial. A user study \cite{cui2024effects} performed in the context of practical software development evaluated the impact of generative AI on developers' productivity involving around 5,000 developers at Microsoft, Accenture and a Fortune 100 company. Their results show a 26\% increase in completed tasks among developers using AI tools, with less experienced developers showing higher adoption rates and greater productivity gains. Similarly, \cite{gonccalves2024assessment} studied a team of eight developers working on diverse projects over a period of twelve weeks. The study suggested a positive value of Copilot (a code generation tool powered by the Codex model) in terms of monetization --- the company is willing to maintain the tool or pay to continue using it. These studies demonstrate the potential of LLMs to improve software development efficiency in practice.

Several human studies have explored the usability of Copilot in programming education. Prather et al. \cite{prather2023s} observed introductory-level students using Copilot for a CS1 assignment. While most students felt Copilot helped them write code faster, they expressed concerns about not fully understanding the generated code and becoming overly dependent on the tool. Similarly, Mailach et al. \cite{mailach2025ok} studied CS2 students using a chatbot to solve programming tasks and found that those using the chatbot achieved 21\% higher scores on implementation tasks. 
\cite{scholl2024novice} assessed ChatGPT-3.5 for solving programming tasks by novice programmers in higher education, finding that while students widely adopt GenAI, their engagement varies from passive acceptance of solutions to critical use. 
\cite{kazemitabaar2023novices} studied 33 learners aged 10–17 and identified common usage patterns of AI code generators.
Unlike the present study, which considers a variety range of participant experience levels, those studies specifically explore the experience of interacting with Copilot among programming novices.
In a setting similar to the present study, Wang et al. \cite{wang2024characterizing} examined how users employ LLMs to solve simple coding problems and fix real-world bugs in small-scale open-source projects.  Across these studies, interaction patterns reveal both the perceived usefulness of LLMs and recurring issues such as hallucinations, overconfidence, and overreliance. Notably, none of these user studies involve formal verification; they assess the correctness of generated code only through manual inspection or program execution.

Shein \cite{SheinEducation} asked three groups of MIT students to produce a program in a language they did not know, FORTRAN, respectively with two different search engines and without any (but access to search), a setting not unlike the present study's. LLM-equipped students performed much faster on program production, but the no-LLM group beat them handily when it came to explaining it some time later. Some results of the present study go in the same direction of suggesting that ``LLMs favor the diligent'' (section \ref{conclusions}).

Other research has focused on Copilot’s impact on task performance and developer behavior. Vaithilingam et al. \cite{vaithilingam2022expectation} found that participants using Copilot were less successful in accurately completing tasks compared to those relying on IntelliSense in VS Code, though they appreciated Copilot’s ability to generate useful starting code --- even when it sometimes led to “debugging rabbit holes.” Barke et al. \cite{barke2023grounded} identified two modes of interaction with Copilot: acceleration mode, where developers use it to complete code they already plan to write, and exploration mode, where they rely on Copilot for unfamiliar tasks. However, over-reliance and the need to choose from multiple suggestions were found to cause cognitive overload and hinder task completion. Although certain interaction patterns observed are similar, this present study distinguishes itself by examining the usability of LLMs for a significantly more advanced goal --- fixing formally verified software.

\section{Tool stack} \label{tools}

The experiments are performed online, using a Web browser \cite{experimentwebsite}; participants do not need to download any other software. The code for the examples is in Eiffel. All software processing is performed by the AutoProof program verification system for Eiffel (originally presented in \cite{tschannen2015autoproof}). Participants access AutoProof through an installation at our institution \cite{autoproofsite} (URL anonymized in this reference). In the Web-based version of AutoProof, the user enters a program text; for each of the tasks, the program is pre-filled with a buggy version. The user can then click the Verify button of the interface, causing compilation and formal verification. In the example below, from the AutoProof tutorial (not a task used in the present study), clicking Verify has resulted in three features being correctly verified (green) and two not, as the prover cannot prove that the respective feature ensures a clause of its postcondition and preserves a clause of the class invariant.


\begin{figure}[t]
  \centering
  \includegraphics[scale=.2]{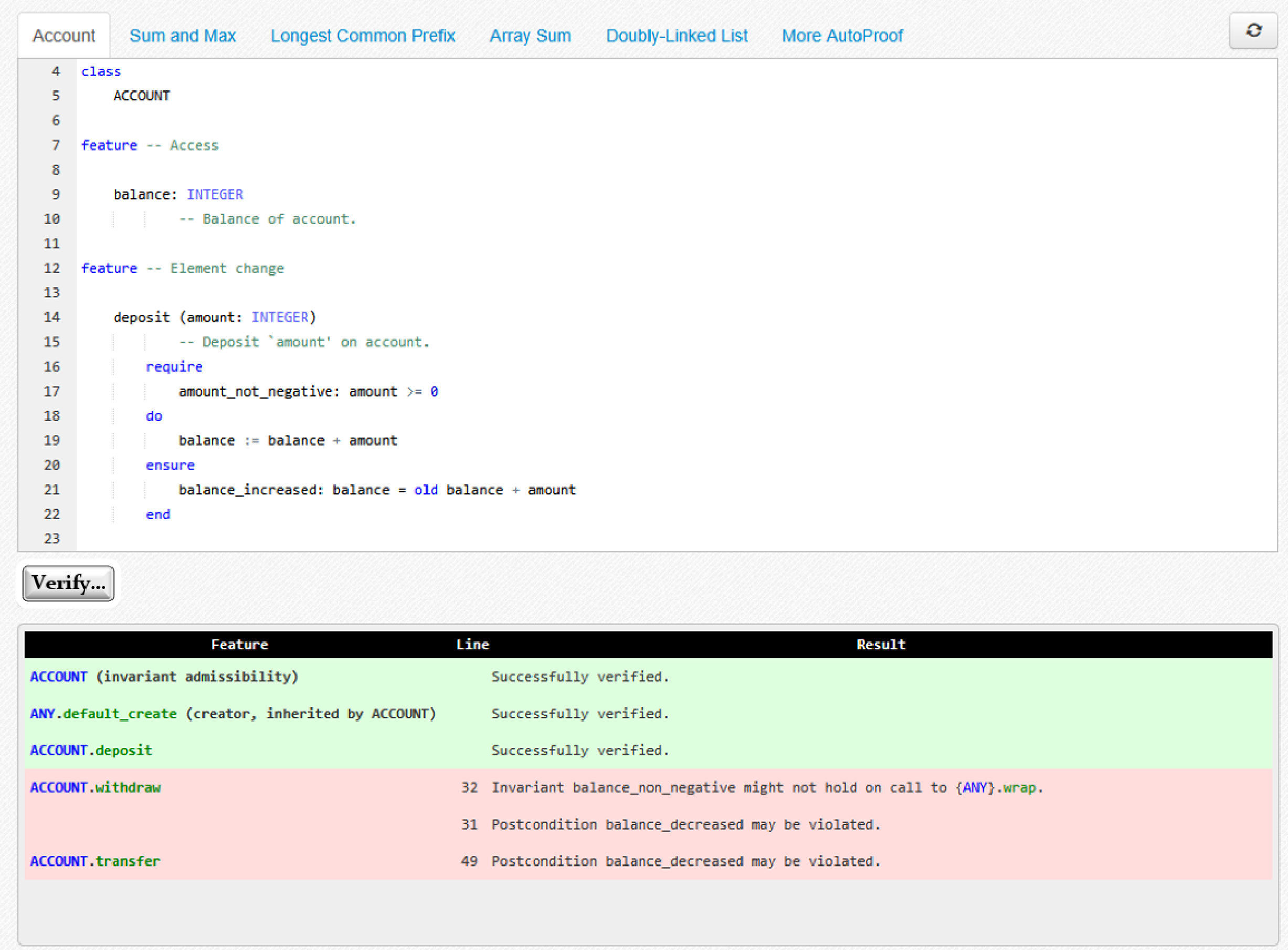}
  \vspace{-0.3cm}
  \caption{An AutoProof session}
  \label{autoproof}
\end{figure}

\noindent AutoProof relies on:
\begin{itemize}
    \item The EiffelStudio compiler for Eiffel. Clicking Verify will first cause a compilation. If the compilation produces an error, AutoProof reports it (and does not attempt any verification). If the compilation succeeds, AutoProof proceeds with verification.
    \item The Boogie prover \cite{leinoboogie, barnett2005boogie}, itself relying on an SMT (Satisfiability Modulo Theory) solver, currently Z3 \cite{de2008z3}. 
\end{itemize}

\noindent For each task in the experiment (see next section), the user is presented with a pre-filled program, which is valid Eiffel but buggy: clicking Verify will result in successful compilation but failed verification, with appropriate diagnostic such as a postcondition not being verified. The user can then correct the program and try again (or give up). 

Although testing tools such as AutoTest \cite{autotest} are otherwise available in the EiffelStudio environment, no testing tool is included in the tool stack for the present work. The verification performed in this work is entirely static: \textbf{no code is ever executed, no test cases are ever needed}. ``Verifying'' a class or one of its features (routines, methods) means submitting them to a proof attempt based on principles of Hoare logic as implemented by the AutoProof-Boogie-Z3 combination.

This approach is made possible by the presence in the Eiffel code of built-in specifications known as \textit{contracts} \cite{meyer1992applying}. In the trivial specification of Fig. \ref{autoproof}, the contract includes a precondition (to deposit a value into a bank account, that value must be positive) and a postcondition (as a result of the deposit operation, the balance will have been increased by that value). Contracts also include class invariants, expressing consistency properties of all instances (objects) of a certain type (class). A program --- possibly resulting from an attempt at correcting a buggy first version, as in the present set of experiments --- passes verification if all creation procedures (constructors) of a class ensure its  class invariant, and every exported feature, executed with its precondition and class invariant satisfied, terminates with its postcondition satisfied and the invariant satisfied again. (This characterization of Hoare-style semantics for object-oriented programs summarizes the principal ideas; see for example the specification of AutoProof \cite{tschannen2015autoproof} for the precise rules.)  

\section {Study setup} \label{setup}

All participants in the study were volunteers, responding to calls for participation sent to students and researchers in our institution, the Eiffel User Group, LinkedIn, and personal contacts. The requirement was simply to have some programming experience, at least basic knowledge of Eiffel (so as not to be derailed by language comprehension issues) and to be willing to devote two hours to the program debugging session, the exact time being chosen by each participant freely, over a 5-day period (which included a week-end for convenience).

The invitation, in its entirety, stated: ``\textit{Dear Colleague, to help with research on program  verification at}'' [name of our organization and research group]\textit{, we are inviting to take part in a test next week. It takes place in a 2-hour slot that you can pick, within those days (including the week-end) to match your schedule. The task is to consider a few buggy Eiffel programs and to attempt to correct the bugs. If you are willing to participate, please confirm by email to} [email address of the contact].'' There was no other element; in particular there was no mention of artificial intelligence, LLMs etc. --- only ``\textit{program verification}''. Assignment to a group mentioned "Group 1" or "Group 2", with no reference to AI or no-AI. The same discipline of not revealing the aims of the study, intended to avoid any bias, was observed in subsequent interactions with participants prior to their sessions; for example the email with a link to instructions was simply titled ``Program debugging study''.

Participants were divided into two groups. Given the relatively small size, we did not try to balance them explicitly by participant characteristics but made a random assignment. The instruction sheet differed between the groups: one prescribed the use of AI tools, the other proscribed the use of any tools. The instruction sheet templates, with identifying URLs removed, are available as part of the supplementary material of this article \cite{experimentinstructions}.

Each instruction sheet was actually generated differently for each participant (of either group) since it included a link to a video session (MS Teams or Zoom) unique to the participant. The instruction sheet specifies the following steps (elements in \textit{italics} are from the instructions, others are comments for this article):
\begin{itemize}
    \item \textit{Fill a short preliminary questionnaire.} The text of the questionnaire remains available \cite{questionnaire2025}.
    
    \item \textit{Connect to a video session at} [URL of the participant's unique Teams or Zoom session]. \textit{Make sure to enable screen sharing }(for our detailed analysis).

    \item \textit{In your personal page, you will find a list of links to specific tasks.} Each task is a link to an AutoProof session, pre-filled with program text (as in Fig. \ref{autoproof}).
    
    \item \textit{For each task, click Verify}. Since the program is buggy, verification will fail with error messages. 
    
    \item (For each task) \textit{attempt to correct the bug, and for each correction attempt click Verify to check the correctness of your proposed solution}. Here the instruction sheet, for both groups, included:  \textit{You are  allowed access to the AutoProof tool tutorial (including examples of fixes) at} [URL of the tutorial at our institution.] \textit{ No search engines or any other docs or websites are allowed (with or without AI).} But then they differed:
    \begin{itemize}
        \item For group 1 (AI): ``\textit{You are allowed interaction with the default model (GPT-4o mini) at} \href{https://duck.ai}{https://duck.ai}.''

        \item For group 2 (no-AI): \textit{You are not allowed interaction with any LLM (Large Language Model) or any other AI tool.}
    \end{itemize}

    \item \textit{After each task, send your solution (working or not) to }[email address] \textit{using the “Share Code URL” button on the AutoProof page and send the generated link.}

    \item \textit{Stop after two hours at most.}
\end{itemize}
\noindent Examination of the results indicates that the instructions were clear and that the participants complied, although one (in the non-AI group) forgot to share the screen, but still delivered the corrected programs. Participants were offered no incentive to perform in any particular way (in particular, student participants' performance had no influence on their grade in any course); the only motivation, expressed in the invitation email reproduced above, was to ``\textit{help research on program verification}''. Examination of the results suggests that all participants took the tasks seriously and did their best to perform at their best.

The results available to the researchers, on which this article is based, include  for each participant: the filled questionnaire; the solutions (corrected programs); the full video (usually two hours) of the session (except for one participant as mentioned. The benefit of the videos is that they allow us to follow the participants' successive attempts and ideas, successful or not. The downside is that this analysis is labor-intensive, since altogether it required watching a total of some 50 hours of recording. Although aided by some custom-produced scripts (allowing an analyst to observe specific events that occur at a specific time), the process is manual.

The questionnaire \cite{questionnaire2025} is short and intended to help analyze the results along different categories. The questions addressed: age (which would have been relevant with a more complete version of query QH6, but is ignored in the present work in favor of experience); number of hours spent weekly on, respectively, programming (in general),  Eiffel programming and LLM usage (ChatGPT, Gemini etc.); duration in years of, respectively, programming experience, Eiffel programming experience and static analysis experience; number of hours spent weekly on Eiffel programming; highest educational degree. The questions were single-choice, from a set of predefined possibilities; their goal was mainly to allow the researchers to get an idea of the level of experience of the participants, typically grouped into two categories (experienced/novice in Eiffel etc.) for the results of section \ref{analysis}.

While the participants' identity is known to the researchers, their anonymity is guaranteed for anyone else (including between the participants themselves) and the researchers are pledged not to reveal any information beyond what is contained in this article. Both the article and its supporting material have been cleaned of any element that could in any way serve to disclose such information.  

29 people volunteered, which we split into a group of 15 (non-AI) and one of 14 (AI). 25 of the volunteers actually performed the experiment. By a stroke of back luck the four no-shows were all from the same group (the AI group), causing a small imbalance. In all we have 10 exploitable AI-assisted results and 15 non-AI-assisted (including the one without video). 

\vspace{-0.3cm}
\section{Threats to validity} \label{threats}

The preceding description of the setup sets the limits of what one can expect from the results. The main limitation is the relatively small number of participants, although it is comparable to the size in previous studies of similar topics (such as \cite{barke2023grounded}). This article consequently refrains from any detailed statistical analysis (which the numbers do not justify) and from drawing sweeping general conclusions.

There was no particular a priori thesis among the researchers. The motivation for the study was the observation, in some of the authors' prior work on Automatic Program Repair, involving sophisticated techniques of repair suggestion, that sometimes ``just asking ChatGPT'' would yield what seemed like similar answers. We wanted to explore whether that impression was borne out by the facts. We can find solace (and disappointment) in either the result that LLMs do very well or that they do not. We just wanted to find out.

As noted in the previous section, all participants took the tasks seriously and did their best. One feature of the experiment that may raise questions is that participants are known to researchers since they share their screens during the session. It is possible that some participants in the AI-assisted group wanted to show (for reputation's sake) that they could solve the problems without AI help. We have, however, no evidence that such a phenomenon took place at all. Being able to observer the participants' actual programming attempts in the recorded (through screen-sharing) sessions, while a tedious effort for the researchers, yielded invaluable information which would not have been available had we attempted full anonymity (without any guarantee of achieving it). 

All participants in the LLM group used only one LLM, GPT-4o mini from Open AI, through the \href{https://duck.ai}{duck.ai} interface. The study does not provide insights into how participants would have fared with another LLM. We simply note that ChatGPT, of which GPT-4o is part, is the dominant LLM offering and often serves as the reference in this field.

\vspace{-0.3cm}

\section {Analysis of the study's outcome} \label{analysis}

This section examines the questions and metrics introduced in section \ref{questions} as ways to address the research questions W and H. The metrics should be interpreted with caution given the small size of the samples, but we believe that they do identify relevant trends. (The next two sections provide more  observations.) The results are examined in the order of the research questions and the corresponding queries and metrics introduced in section \ref{questions}. We provide the raw results when relevant, followed by our observations, particularly emphasizing outcomes that may go against expectations.

To enable more significant visual inferences from the results comparing participants' performance on various tasks, such as QW-2, QW-3 etc., the charts show these tasks in order, from left to right, of increasing difficulty  (as assessed by us from observing the participants' performance).

The  diagrams use: {\color{blue}blue} for the group that used AI and  {\color{orange}orange} for the group that did not; `$\bullet$' for the average; and `$\times$' for the median; $\top$ and $\bot$ for the 80th and 20th percentile respectively. If a data point does not exist it is omitted in the diagram; for example there is no blue mark (in the diagram for QW-2 below) for Task 9 since no one using AI solved it successfully.

There are 9 tasks altogether, each consisting of an Eiffel class with a buggy feature (method, routine). The source code, as well as the corrections for the bugs, are available  at \cite{experimentbugs}. The tasks are of  varying complexity and difficulty levels; they are drawn from several sources including  common examples from formal verification tutorials, bug databases, and examples from EiffelBase libraries that at some point of their existence contained actual bugs, corrected since.
The order in which the tasks appear in the result charts below, slightly different from the order in which they were presented to participants, is the increasing order of their difficulty (based on our assessment, confirmed by observation of the participants' performance). Each corresponds to one of the problems in \cite{experimentbugs}:
\begin{itemize}
    \item Task 1. \e{MAPLE\_RECURSIVE\_ABSOLUTE}: The class defines an \e{abs} function intended to return the absolute value of an integer\e{x}. The function incorrectly returns \e{x} instead of \e{-x}, failing to handle negative inputs properly.
 \item Task 2. \e{FIND_IN_SORTED}: The class implements a recursive binary search to locate an integer in a sorted array, returning its index or 0 if not found. Injected bug: an incorrect conditional that prevents the function from correctly identifying found elements.
 \item  Task 3. \e{CALCULATOR}: The class defines a calculate function that performs basic arithmetic operations (+, -, *, /, \%\%) on two integers based on the given operator, returning -1 for unsupported operations. Injected bug: in the implementation of the subtraction case, the code incorrectly performs addition instead of subtraction.
 \item  Task 4. \e{LINKED_STACK_MAKE_COMBINED}: The class represents a linked stack structure with state-tracking booleans (\e{before}, \e{after}) and pointers to elements, and includes a creation procedure to initialize the stack. Bug: the creation procedure incorrectly sets \e{before := True} without properly initializing other attributes (e.g., \e{first_element} and \e{last_element}).
 \item Task 5. \e{TIME}: The class models a 24-hour clock with operations to set and retrieve time components, decrement time, run countdown timers, compute time differences, while maintaining strict invariants on valid time values. Injected bug: the precondition of the routine \e{set_hour} incorrectly allows \e{hour} to take the value 24, violating the class invariant on the bounds on \e{hour} (\e{0 $\leq$ hour} \e{and hour $<$ 24}).
 \item Task 6. \e{QS_QUEUE}: the class implements a fixed-size queue (max size 100) with operations for inserting, deleting, searching, and checking elements, while maintaining front and rear indices and tracking overflow/underflow exceptions. Bug: in the \e{search} routine, the loop incorrectly decrements \e{index} (\e{index := index - 1}) instead of incrementing it, causing an infinite loop or incorrect behavior when the key is not found.
 \item  Task 7. \e{PRIME_CHECK}: the class provides a routine \e{is_prime} that checks if a given integer \e{a} is prime by testing divisibility from 2 up to \e{a // 2}. Bug: incorrect exit condition --- when a $\backslash \backslash$ i = 0 (a divisor is found), Result is set to False but the loop does not break.
 \item  Task 8. \e{ARRAY_FORCE_TO_EMPTY}: the class  implements an indexable container with arbitrary bounds and contiguous memory storage, including a buggy routine intended to insert an element into an empty array at index \e{i}: when the array is resized, the last element uninitialized (wrong default handling).
 \item  Task 9. \e{FIND_FIRST_IN_SORTED} class defines a routine that attempts to find the first occurrence of key in a sorted array using a binary search strategy but contains logic errors: when the $key$ is less than the \e{middle} element, it fails to shrink the search interval properly, potentially causing an infinite loop or incorrect result.
\end{itemize}
\vspace{-0.4cm}

\subsection{Goal W: In turning a buggy program into a correct one, is it fruitful to use an LLM?} \label{goalW}

\begin{itemize}
    \item QW-1 Can programmers solve more debugging tasks with/without LLM?
          \begin{itemize}
              \item [] MW-1.a How many tasks are solved with LLM?
              \item [] MW-1.b How many of the tasks are solved without LLM?
          \end{itemize}


    \includegraphics[scale=.5]{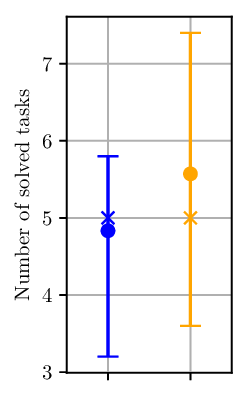}

   \noindent \textit{Observations}: The result is a rough measure; subsequent charts provide more detailed views,  refined along various criteria. This one provides, however, the first surprise: we would, perhaps naïvely, expected the AI group to fare better overall. Here all indications --- minima, maxima, medians --- are in the reverse order. A simplistic deduction from looking at just that figure would be ``LLM help for debugging hurts rather than helps''. The results are too coarse-grained to allow such a conclusion but they point the study in an interesting direction.

   Variation is wider in the non-AI group, presumably (a conjecture to be assessed better in results appearing below) because of the varying level of expertise.

    \item QW-2 For tasks that programmers can solve with/without LLM, which takes longer?
          \begin{itemize}
              \item [] MW-2.a How long does each task solved with LLM take?
              \item [] MW-2.b How long does each task solved without LLM take?
          \end{itemize}
                    \includegraphics[scale=.5] {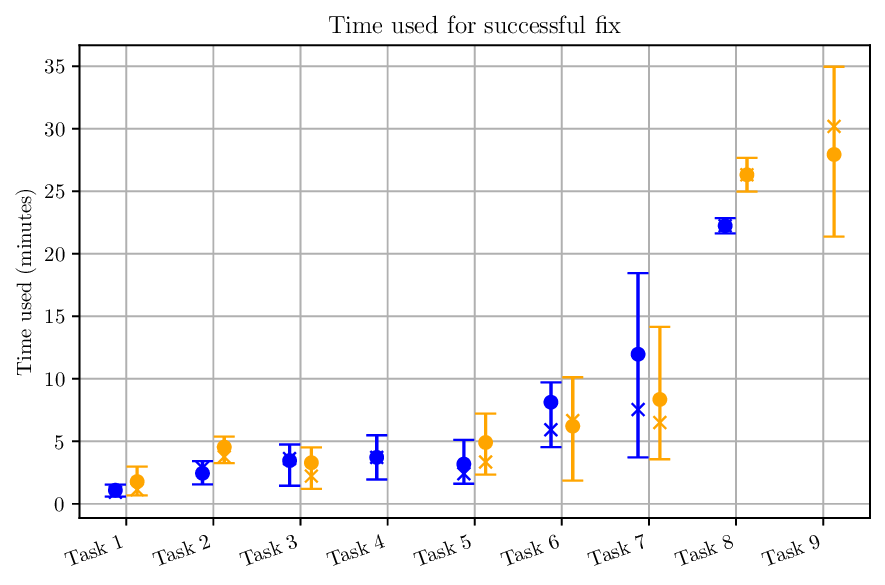}

    \noindent \textit{Observations}: The information on time spent is not very conclusive but suggests that LLM help is somewhat effective for the easier tasks. For the harder tasks, LLMs helps for Task 8, but then no LLM-helped participants finished Task 9, the hardest.

    \item QW-3 For tasks that programmers cannot solve with/without LLM, which takes longer?
          \begin{itemize}
              \item [] MW-3.a How long does each task unsolved with LLM take?
              \item [] MW-3.b How long does each task unsolved without LLM take?
          \end{itemize}

   \noindent 
   \noindent \textit{Observations} (the chart appears at the top of the next column): Here a worrying phenomenon seems to affect hopes of using LLMs for correcting programs: spending far too much time in a useless direction. In observing participant sessions, we noticed a number of ``hallucination loops'', where the LLM was giving wrong advice and the participant got hooked with no prospect of success. Section \ref{antipatterns} discusses this phenomenon further.

         \includegraphics[scale=.5]{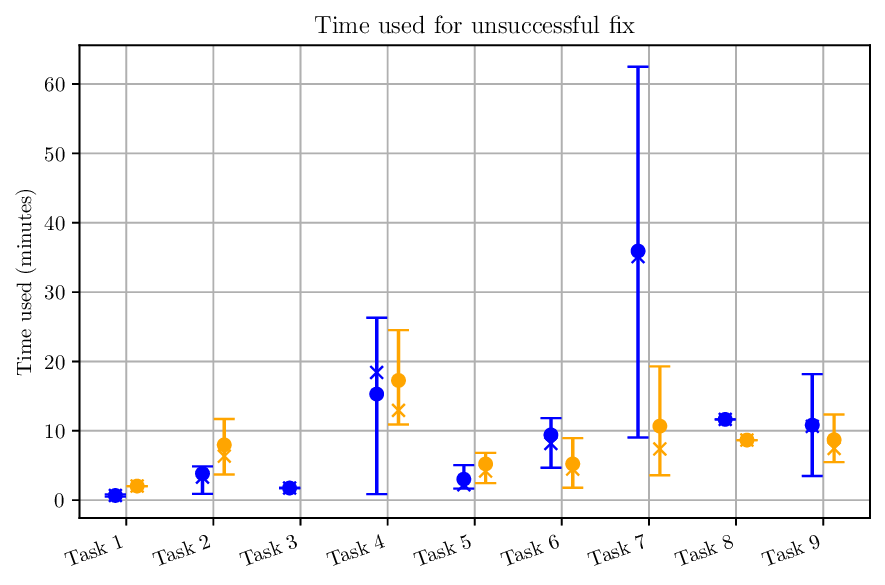}

    \item QW-4 Do submissions contain more incorrect features with/without LLM?
          \begin{itemize}
              \item [] MW-4.a How many tasks are incorrectly solved with LLM?
              \item [] MW-4.b How many tasks are incorrectly solved without LLM?
              \item [] MW-4.c How many tasks are correctly solved with LLM?
              \item [] MW-4.d How many tasks are correctly solved without LLM?
          \end{itemize}
          \includegraphics[scale=0.5]{figures/metrics/Solved_tasks_ai_vs_non_ai.eps}
          \includegraphics[scale=0.5]{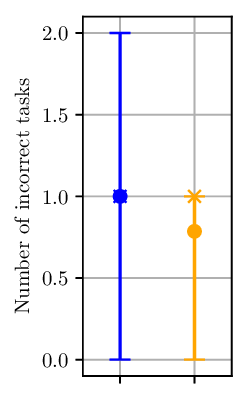}

    \noindent \textit{Observations}: The data is not great advertisement for AI support for debugging. Non-AI-assisted programmers solve more tasks correctly and fewer tasks incorrectly!
     
    \item QW-5 Do LLMs help experienced programmers more/less than novice ones?
          \begin{itemize}
              \item [] MW-5.a How many problems do experienced programmers
              solve with/without LLM?
              \item [] MW-5.b How many problems do programmers with static verification experience
              solve with/without LLM?
              \item [] MW-5.c How many problems do programmers with programming language experience
              solve with/without LLM?
          \end{itemize}
          \includegraphics[scale=0.5]{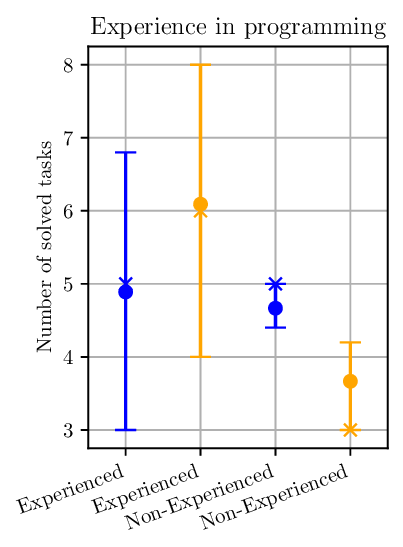}
          \includegraphics[scale=0.5]{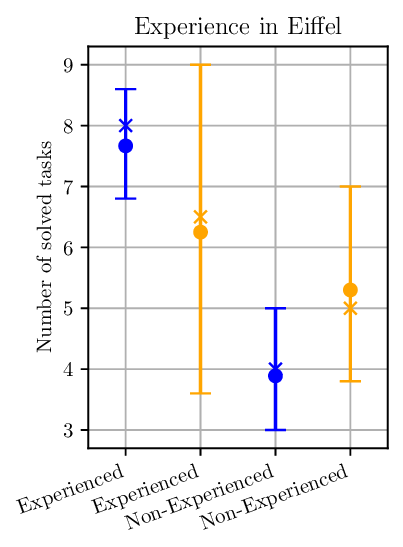}\\
          There were too few people experienced in static verification to draw a meaningful diagram for the associated measures.

    \noindent \textit{Observations}: As noted above, the questionnaire assessed experience in several ranges (0-2, 3-5, 6-10, over 10 years), which we reduced to two in each case (5 years or more being considered ``experienced''). For overall experience in programming, the LLM-help advantage is blatant for novices (right two segments of the left box); for experienced programmers, the situation is reversed, with non-LLM having a smaller but still clear advantage. For experience in the programming language, the picture is less clear; given the wide spectrum of non-LLM results it is difficult to draw a firm conclusion, but it does seem that programmers with a strong experience in Eiffel (leftmost segment in the right box) can benefit most from LLM use.  We may conjecture that, knowing the language very well --- in addition to their general knowledge of programming, which is implied by their specific experience in the language --- they have a long practice of sifting through code in that language and correcting bugs, enabling them to use an LLM suggestion much more effectively than a beginner through their ability to sort the weed from the chaff (the good from the bad and ugly) by developing a strong hallucination-spotting and hallucination-avoiding instinct. 
    
    \item QW-6 Do LLMs help active programmers more/less than occasional ones?
          \begin{itemize}
              \item [] MW-6.a How many problems do active programmers solve with/without LLM?
              \item [] MW-6.b How many problems do programmers with active static verification practice solve with/without LLM?
              \item [] MW-6.c How many problems do active programmers in the given programming language solve with/without LLM?
          \end{itemize}
          There were too few people that practiced Eiffel or static verification regularly to draw meaningful charts.
\end{itemize}

\subsection{Goal H: If programmers do use an LLM for debugging, what is an effective process?} \label{goalH}

\begin{itemize}
    \item QH-1 What are the categories of the prompts used by the programmers?
    (clarification, expansion of previous, skepticism of previous answer, other categories use in the related works)
          \begin{itemize}
              \item [] MH-1.a In using LLMs for debugging, what are the categories of prompts (determined empirically from the evidence and from previous published work)?
              \\
              We identified the following categories, given here with their percentage of use in the observed results (from watching the recorded screen-shared sessions):
              
              \begin{itemize}
                  \item Request for context clarification: 19.5\%.
                  \item Persona (a frequent recommendation in texts about ``prompt engineering'': tell the LLM that you are asking a question in a certain role (as in ``\textit{assume that I am a novice Eiffel programmer. How would I...}''): 2.3\%.
                  \item Other categories cited in the literature: not present in the experiment's record.
              \end{itemize}
              \item [] MH-1.b For each task, how many categories of prompts do programmers use?
              \textit{Note:} There is too little data. Categories were not used.
          \end{itemize}

           \noindent \textit{Observations}: The overall conclusion here is that programmers practice no prompt engineering beyond requesting clarifications. 

    \item QH-2 What are the components of the prompts for solved/unsolved submission?
          \begin{itemize}
              \item [] MH-2.a How many prompts contain natural language? (Answer: 97.7\%.)
              \item [] MH-2.b How many prompts contain code? (Answer: 48.3\%.)
              \item [] MH-2.c How many prompts contain tool output? (Answer: 43.7\%.)
          \end{itemize}
    \noindent \textit{Observations}: The total is more than 100\% since a prompt can contain elements of several kinds. Natural language is essentially always present. Code is present half of the time; so is the output of AutoProof or EiffelStudio, which suggests that programmers like  to feed some of the messages they get in order to get feedback and explanations from the LLM. 
    
    \item QH-3 How important is the phrasing of prompts in solving a bug?
          \begin{itemize}
              \item [] For each category of prompts, how many prompts lead to a solved task?
              \item [] MH-3.b How many prompt categories lead to a solved task?
          \end{itemize}
          \textit{Observations}: There was too little data to answer this query meaningfully.

    \item QH-4 How do programmers interact with LLMs?
          \begin{itemize}
              \item [] How many prompts do programmers send before solving a task? (Answer: 2.8 prompts on average.)
              \item [] How many prompts do programmers send for unsolved tasks? (Answer: 3.6 prompts on average).
          \end{itemize}
    \noindent \textit{Observations}: These figures suggest that the interaction with the LLM is meaningful and potentially useful. They also indicate that LLMs are not genius debuggers: even for solving a task, you need almost three steps on average; and before you realize that the LLM does \textit{not} help you have interacted between 3 and 4 times.
    
    \item QH-5 How do programmers use the outputs from LLMs?
          \begin{itemize}
              \item [] How many times do programmers copy-paste a fix produced by LLMs?
              \item [] MH-5.b How many times do programmers accept LLMs’ output as the final version?
          \end{itemize}
  
    \noindent \textit{Observations}: We do not have the exact numbers for the given questions but we do have interesting information from watching the recorded sessions. Half of the participants \textit{did not copy-paste at all} from LLM answers. Those who did copy-paste did so \textit{less than once per task} (except one person who copy and pasted 18 times). Only once was the copy-pasted result part of the final version. We may safely deduce that in this experiment participants do not slavishly use the LLM to give them a solution, but as guidance; they still exercise their own judgment.

    \item QH-6 What is the effect of programmers’ experience on LLM use?
          \begin{itemize}
              \item [] MH-6.a Do experienced programmers use more/fewer prompts? (Answer: experienced programmers use 4.9 prompts on average, median 4; non-experienced programmers use 25 on average, median 23.)
              \item [] MH-6.b Do experienced programmers use LLM output more/less often? (Answer: experienced programmers use on average 0.9 times the output, median 0; non-experienced programmers use 10.0 times the output, median 7.) 
          \end{itemize}
\end{itemize}

\noindent \textit{Observations}: These results confirm that experienced programmers, when they use the LLM, get much faster to the point, presumably because they understand the underlying programming issues much better and can identify useful advice quickly, removing irrelevant or incorrect elements.  

\section {Qualitative observations} \label{observations}

Analysis of participants' behavior, in addition to their actual code outputs,  yielded insights not captured by the mostly quantitative results of the preceding section. This further analysis is made possible by examination of the 24 individual recordings of participants' shared-screen sessions.



\subsection {Antipatterns of LLM usage} \label{antipatterns}

A real risk observed in a number of sessions is the \textbf{hallucination loop}. The AI tool makes an incorrect guess about how to correct a bug; the programmer tries it, the result does not verify, the programmer tries to fix it, and gets into a fruitless try-verify-fail loop. The problem is that a fix might look plausible but, in the current state of generative AI, have no conceptual basis. A typical scenario occurs for the most challenging tasks (8 and 9) of the present study, for which the LLM was unable to produce a correctly fix. Verification of its proposed  code failed; usually, the participant still failed to understand the cause after several iterations.

Another aberrant behavior, where using the LLM actually harms the programmer, is a variant of the hallucination loop, the \textbf{noisy solution}. The code proposed by the LLM actually contains the solution, but also includes, along with the correct elements, irrelevant ones which prevent or delay the programmer from obtaining a correct fix. Copy-pasting the solution will not work. 
An example of such behavior occurred in Task 9: the participant provided the buggy code to the LLM, which reported multiple issues (some of them are irrelevant or spurious) and suggested several possible fixes; the participant tried out these suggestions one by one but, before reaching the correct one, became frustrated and gave up.  

More generally, the LLM often exhibits the \textbf{neutrality} flaw: it presents various solutions, some potentially useful, others  worthless or harmful, without indicating their respective likelihood of being helpful. In addition to pointing to an area of clearly needed improvement for LLMs, this phenomenon suggests that programmers using them should not just treat all their suggestions as equally valid but make a systematic effort to \textit{rank} them (see \ref{advice} below).

Watching the videos uncovers yet another antipattern: \textbf{timidity}. Some programmers, seeing that an LLM-generated fix is not  working after a copyppaste, tried their luck by submitting to the verifier small variations of the AI solution without following a rigorously logical process. Poorly thought-out use of AI tools may lead to this unproductive behavior.

\subsection {Patterns of AI usage, and their effectiveness} \label{patterns}

Our observation of how participants interact with LLMs yield the following categorization of programmer personalities. Our data shows that all categories arise in practice (a larger study would make it possible to determine reliable percentages of their occurrence). They are the following in --- roughly --- order of increasing use of the AI tool (and increasing sophistication of that use). 

\begin{itemize}

    \item The \textbf{independent}: Does not use AI support even when available.

    \item The \textbf{prejudiced}: Frames prompt around a prior hypothesis, such as``it is the contract that is buggy'', steering the AI tool off-course.

    \item The \textbf{searcher}: Uses LLM like a reference, encyclopedia or search engine. Asks, for example, for explanations or reminders on language mechanisms, such as \textit{“What does} [the keyword] \textbf{detachable} \textit{mean in Eiffel?}”)\footnote{In the void-safety mechanism of Eiffel, which guarantees the absence of null-pointer responsibility as part of the type system, a type is by default ``attached'', guaranteeing that its values will never be null (void); in the other case, that of a variable can have a null value and hence must be protected, the type must be marked \textbf{detachable}.}.The searcher remains in control and issues no queries requesting fixes.

     \item The \textbf{copy-paster}: Queries the AI tool for fixes in the form of code. Pastes responses back with little interpretation or adaptation. Prone to “hallucination loops” when verification fails.

     \item The \textbf{follower}: Accepts the AI tool's suggestions and tries them out one-by-one. Unlike the \textit{copy-paster}, works from these suggestions rather than taking them unquestioningly.

    \item The \textbf{collaborator}: Uses the AI tool as a form of pair debugger. Supplies code and/or tool output, interprets LLM suggestions and decides to apply them or not. Adapts the generated code from LLM; progressively enrich prompts. Our observations consistently indicate that this personality pattern is the most successful. It serves as the basis for our proposed ideal strategy (\ref{advice} below).
 
\end{itemize}

\noindent It might intuitively seem that individuals will oscillate between these attitudes, but that does not occur in our observations (with one exception seen next: turning into a quitter). Throughout the experiments, each individual stuck to one of the patterns.

\textit{Collaborators}  had the highest observed success rate (\~91\%), especially when the verifier failed to prove the proposed fix and provided a counterexamples from failed  outputs, enabling them to apply selective edits rather than pasting wholesale. \textit{Copy-paster} behavior was common (38\%) but low-yield (only \~33\% helpful) and associated with cascading compilation failures when the LLM rewrote large code blocks. \textit{Followers} and \textit{Prejudiced} had near-zero direct success and had a high chance of turning into \textit{quitters}, typically after two negative high-effort LLM sessions.

\subsection{Advice for LLM usage} \label{advice}

In addition to categories of LLM usage for debugging and repair, the extensive observation of participants' tackling of the problems, their attempts, their successes and their failures yields the following suggestions for a winning strategy of taking advantage of AI for these tasks.

\textbf{Smell a rat}: Detect strong user framing (as in ``\textit{the bug must be in the contract}''), appealing in their simplicity but useless, and look at alternative hypotheses. In particular, be on the lookout for incorrect early hypotheses from the LLM, into which the \textit{Prejudiced} personality would jump head-on. They are the primary causes of frustration in LLM usage, observed again and again in our experiments.

\textbf{Remove noise}: Make your way through multi-issue early answers in which the correct fix is present, but buried deep. The LLM does not separate the essential from the auxiliary; you should.

\textbf{Signal highlighting}: When you see suggestions, rank them (to fight the ``neutrality'' antipattern from  \ref{antipatterns}). Visually highlight one-line high-confidence fixes so that they are not buried.

\textbf{Too good to be true}: Avoid the temptation to copy-paste the full solution (as the \textit{copy-paster} would do), since it is likely to lead to high failure rate, often beginning with a compilation failure (the LLM is not an expert in the programming language).


From the present work also emerges a proposed \textbf{best overall strategy}, based on the \textit{collaborator} personality pattern. where at each step you:
\begin{itemize}
    \item Provide both the candidate code and the \textbf{verifier's output}.
    \item Ask not just for a proposed solution but for the ``\textbf{diff}'' (list of individual differences) with the previous version.
    \item \textbf{Review} the proposed fix. (This rule is a general one: only make sure you understand a proposal before submitting it.)
\end{itemize}
Throughout this process, the programmer should constantly be on the alert for the risk of taking the wrong road (by following misplaced LLM advice). The old advice, ``if you find yourself in a hole, stop digging'', applies. One should know when to stop fiddling with an unverified answer in the hope that one more twist will make it verify. Sometimes the best strategy is to start afresh, with a mix of LLM advice and personal thinking.

\section {Conclusions} \label{conclusions}

Section \ref{threats} stated the limitations of this study, particularly regarding sample size. Counterbalancing them is a degree of certainty not found in many Automatic Program Repair studies which rely on tests: here, through the use of a mathematically-based and tool-supported environment for proving program correctness, we can determine unambiguously that a proposed fix is correct (in the specific sense of satisfying the stated specification, and within the limitation of proof tools) or not.

\vspace{-0.15cm} 

\subsection{Lessons on the use of LLMs for debugging} \label{lessons} 

One can derive a few lessons from the experiment. One stands out clearly: whatever their merits for debugging, LLMs (at least to judge by GPT-4o mini's performance) are not a silver bullet. In fact programmers not using the LLM perform better or as well on most counts. Two exceptions are, at opposite ends: complete novices, who can benefit from the LLM to identify and correct \textit{simple} bugs; and programmers who are experts in the language, who would probably have found the corrections by themselves but can use the LLM to zoom in faster on the solution, saving some time. 

The study also uncovers a major issue with LLM-based debugging: the hallucination loop (section \ref{antipatterns}). Hallucination for LLMs is well-known, but in program debugging we encounter a particularly disruptive form: a suggestion that has all the looks of reasonableness, but misleads the programmer into a fruitless direction. Often, expert programmers can smell a rat and ignore hallucinating answers, but novices are at a serious risk of failure.

\vspace{-0.15cm} 

\subsection{Contributions} \label{contributions} 

We believe the present study, taking account of the numerical limitations of the experiment, provides an important perspective and some clear results which we hope will inform and help others who may wish to pursue studies with a larger programmer sample enabling statistical inferences. The contributions include:

\begin{itemize}
    \item The use of a \textbf{program prover} for conducting software engineering studies, in the present case the assessment of Automatic Program Repair strategies, to provide an incontrovertible answer as to the correctness or incorrectness of the results --- as opposed to the dominant use of tests (or, sometimes, manual inspection), from which any correctness conclusion is inevitably shaky.
     
    \item The \textbf{methodology}. The GQM-based set of goals, resulting queries and supporting metrics, devised before the experiment, proved adequate for it without change, and capture, we believe, what should be studied when assessing the value of AI support for debugging and automatic program repair.
    
    \item The use of full-session \textbf{screen-sharing recordings}. An argument against this approach is the waiving of programmer anonymity vis-à-vis the researchers (although anonymity is preserved for the world at large in the results of this study). In addition, the approach is labor-intensive, since someone must examine the videos\footnote{It is not clear whether today's AI tools can help in this analysis.}; with 25 participants it implied scrutinizing 50 hours, and the approach may be hard to sustain for a study that would include a few hundred programmers. But for the present study the ability to observe how programmers work, step by step, proved immensely valuable and yielded insights --- on the use of prompts and the use of copy-paste ---  that we would never have obtained by just looking at raw code produced by the programmers.
    
    \item The identification of \textbf{personalities} of LLM usage and their patterns (section \ref{patterns}). 
    
    \item The identification of the risk of \textbf{hallucination loops} through which, particularly for non-experts, LLMs can take a programmer completely off-course and cause more harm than good.
    
    \item The painful realization that the enthusing first impression that an LLM can produce when one first tries to apply it to program repair can be followed by \textbf{disillusion}, as the proposed fixes may have only the \textit{appearance} of correctness.
    
    \item A delineation of the \textbf{cases} in which LLM can (nevertheless) help.
    
    \item For these cases, \textbf{validated advice} on how to gain these benefits, including an effective overall strategy
  
\end{itemize}

\noindent The overall lesson is that just as luck favors the prepared, \textbf{AI helps the diligent}. Hoping that feeding a buggy program into an LLM will yield a correctly repaired version seems, in the current state of generative AI, unfounded. As also evidenced by Shein's MIT study of program construction \cite{SheinEducation}, there is no substitute for thinking; at every step you must remain on top of the process and understand what is being presented to you.

If that is the case, software developers can take advantage of LLM debugging help to move more quickly through the issues, separating the essential from the auxiliary, and develop an effective AI-aided process of program repair. 

\noindent \textit{Acknowledgements}. We are grateful to the following colleagues for their help: Eric Bezault, Joel Gamonez, Alejandro García, Karsten Heusser, Jimmy Jack Johnson, Adam Kasprzak, Viktoriya Kananchuk, Alik Khilazhev, Alexander Kogtenkov, Éric Lubat, Vivien Moreau, Anders Persson, João M. Rocha, Vladimir Shcherba, Kseniya Sidorova, Catalin Toma. We are also indebted to Joaquim Soares and Ivan Turlakov of Constructor Institute of Technology for essential help in setting up the debugging sessions and making the results available. After the publication of a first version of this article on arXiv, a fruitful discussion took place on LinkedIn (\url{https://bit.ly/4m30G5F}); we are grateful in particular to Hillel Wayne for pointing out an inconsistency, now corrected, in the terminology. 

\bibliographystyle{splncs04}
\bibliography{reference}

\end{document}